\documentclass{midl} % Include author names

\usepackage{mwe} % to get dummy images

\graphicspath{ {figures/} }
\usepackage{titlesec}

\usepackage{caption}
\jmlrproceedings{MIDL}{Medical Imaging with Deep Learning}
\jmlrpages{}
\jmlryear{2024}

\jmlrworkshop{Short Paper -- MIDL 2024 submission}
\jmlrvolume{-- Under Review}
\editors{Under Review for MIDL 2024}

\title[MR Series Classification]{Automatic classification of prostate MR series type using image content and metadata} 

\midlauthor{\Name{Deepa Krishnaswamy\nametag{$^{1}$}} \Email{dkrishnaswamy@bwh.harvard.edu}\\
\Name{Bálint Kovács\nametag{$^{2}$}} \\
\Name{Stefan Denner\nametag{$^{2}$}} \\
\Name{Steve Pieper\nametag{$^{3}$}} \\
\Name{David Clunie\nametag{$^{4}$}} \\
\Name{Christopher P. Bridge\nametag{$^{5}$}} \\
\Name{Tina Kapur\nametag{$^{1}$}} \\
\Name{Klaus H. Maier-Hein\nametag{$^{2}$}}\\
\Name{Andrey Fedorov\nametag{$^{1}$}}\\
\addr $^{1}$ Brigham and Women's Hospital, Boston, MA, USA\\
\addr $^{2}$ Division of Medical Image Computing, German Cancer Research Center, Heidelberg, Germany\\
\addr $^{3}$ Isomics, Cambridge, MA, USA\\
\addr $^{4}$ PixelMed Publishing, Bangor, PA, USA\\
\addr $^{5}$ Massachusetts General Hospital, Boston, MA, USA}

\begin{document}
\maketitle

%%%%%%%%%%%%%%%%%%%%%%%%%%%%%%%%%%%%%%%%%%%%%%%%%%%%%%%%%%%%%%%%%%%%%%%%%%%%%%%%%%%%%%%%%%%%%%%%%%

\begin{abstract}
With the wealth of medical image data, efficient curation is essential. Assigning the sequence type to magnetic resonance images is necessary for scientific studies and artificial intelligence-based analysis. However, incomplete or missing metadata prevents effective automation. We therefore propose a deep-learning method for classification of prostate cancer scanning sequences based on a combination of image data and DICOM metadata. We demonstrate superior results compared to metadata or image data alone, and make our code publicly available at \url{https://github.com/deepakri201/DICOMScanClassification}.

\end{abstract}
% check the word limit for the abstract

%%%%%%%%%%%%%%%%%%%%%%%%%%%%%%%%%%%%%%%%%%%%%%%%%%%%%%%%%%%%%%%%%%%%%%%%%%%%%%%%%%%%%%%%%%%%%%%%%%


%%%%%%%%%%%%%%%%%%%%%%%%%%%%%%%%%%%%%%%%%%%%%%%%%%%%%%%%%%%%%%%%%%%%%%%%%%%%%%%%%%%%%%%%%%%%%%%%%%

\section{Introduction}
To diagnose clinically significant prostate cancer (PCa), multi-parametric magnetic resonance imaging (MRI) scans are acquired, typically following the PI-RADS v2 guidelines \cite{Weinreb2016,Turkbey2019}. However, metadata describing the type of scan (free text DICOM field \textit{SeriesDescription} configured by the operator) is prone to user input errors, institutional conventions, and removal during de-identification. As most machine learning (ML) tasks for segmentation and detection of lesions require specific sequences as input \cite{picai, prostate158}, efficient curation without relying on free text fields or manual input is necessary. Approaches for series classification have been developed based on metadata \cite{Gauriau2020, Cluceru2023}, and convolutional neural networks (CNN) using images \cite{Kasmanoff2023,Salome2023,VanDerVoort2021}. However most approaches do not combine metadata with image data in a learned fashion \cite{Cluceru2023}, and relatively few methods have performed classification of prostate/pelvis sequences \cite{Baumgärtner2023,Helm2024}.

We propose a CNN-based method for classification of prostate MRI scans, where A) we integrate the image data and DICOM metadata (acquisition parameters captured in fields populated by the scanner) in a single CNN, which has not been done yet for scan classification, B) we train and evaluate our methods using entirely publicly available DICOM PCa MRI collections, and C) we compare our method with a random forest approach using the metadata, and a CNN-based approach using the image data.

%%%%%%%%%%%%%%%%%%%%%%%%%%%%%%%%%%%%%%%%%%%%%%%%%%%%%%%%%%%%%%%%%%%%%%%%%%%%%%%%%%%%%%%%%%%%%%%%%%

% \begin{table}[t]
% \centering
% \floatconts
%   {table1}%
%   \tiny
%   {\caption{Collections from Imaging Data Commons and the corresponding number of MR series included for the analysis. *=some or all of the patients were scanned with an endorectal coil, {\textdagger}=multiple manufacturers, {\ddag}=multiple magnetic field strengths }}%
%   {\begin{tabular}{|l|l|l|l|l|l|l|l|l|}
%   \hline
%   \bfseries Dataset & \bfseries{References} & \bfseries T2W & \bfseries DWI & \bfseries ADC & \bfseries DCE & \bfseries{Train} & \bfseries{Val} & \bfseries{Test} \\
%   \hline
%   QIN-Prostate-Repeatability* & \cite{Fedorov2018A} & 30 & 30 & 30 & 30 & \checkmark & \checkmark & \checkmark\\
%   ProstateX{\textdagger} & \cite{Litjens2014, Litjens2017} & 431 & 357 & 356 & 15456 & \checkmark & \checkmark & \checkmark\\
%   Prostate-MRI* & \cite{Choyke2016} & 26 & 52 & -- & 51 & -- & -- & \checkmark \\
%   Prostate-3T{\textdagger} & \cite{Litjens2015} & 64 & -- & -- & -- & -- & -- & \checkmark \\
%   Prostate-Diagnosis* & \cite{Bloch2015} &  93 & -- & -- & -- & -- & -- & \checkmark \\
%   Prostate-MRI-US-Biopsy*{\textdagger}{\ddag} & \cite{Natarajan2013} & 958 & 110 & 1019 & --  & -- & -- & \checkmark\\
%   & \cite{Sonn2013} & & & & & & & \\
%   Prostate-Fused-MRI-Pathology* & \cite{Madabhushi2016} & 46 & 13 & 12 & 102 & -- & -- & \checkmark\\
%   & \cite{Singanamalli2016} & & & & & & & \\
%   \hline
%   \end{tabular}}
% \end{table}

\begin{table}[]
\centering
  \tiny
  {\caption{Collections from Imaging Data Commons and the corresponding number of MR series (patients in parentheses) included for the analysis. ERC = endorectal coil was used, {\textdagger}=multiple manufacturers, {\ddag}=multiple magnetic field strengths. }\label{table1}}%
  {\begin{tabular}{|c|c|c|c|c|c|c|c|c|}
  \hline
  \bfseries Dataset & \bfseries With ERC & \bfseries T2W & \bfseries DWI & \bfseries ADC & \bfseries DCE & \bfseries{Train} & \bfseries{Val} & \bfseries{Test} \\
  \hline
  QIN-Prostate-Repeatability & \checkmark & 30 (15) & 30 (15) & 30 (15) & 30 (15) & \checkmark & \checkmark & \checkmark \\
  \cite{Fedorov2018A} & & & & & & & & \\
  ProstateX{\textdagger} & -- & 431 (346) & 357 (346) & 356 (346) & 15456 (346) & \checkmark & \checkmark & \checkmark \\
  \cite{Litjens2014, Litjens2017} & & & & & & & & \\
  Prostate-MRI & \checkmark & 26 (26) & 52 (26) & -- & 51 (26) & -- & -- & \checkmark \\
  \cite{Choyke2016} & & & & & & & & \\
  Prostate-3T{\textdagger}& -- & 64 (64) & -- & -- & -- & -- & -- & \checkmark \\
  \cite{Litjens2015} & & & & & & & & \\
  Prostate-Diagnosis & \checkmark & 93 (91) & -- & -- & -- & -- & -- & \checkmark \\
  \cite{Bloch2015} & & & & & & & & \\
  Prostate-MRI-US-Biopsy{\textdagger}{\ddag} & \checkmark & 958 (792) & 110 (108) & 1019 (836) & --  & -- & -- & \checkmark \\
  \cite{Natarajan2013} & & & & & & & & \\
  \cite{Sonn2013} & & & & & & & & \\
  Prostate-Fused-MRI-Pathology & \checkmark & 46 (27) & 13 (12) & 12 (12) & 102 (28) & -- & -- & \checkmark \\
  \cite{Singanamalli2016} & & & & & & & & \\
  \cite{Madabhushi2016} & & & & & & & & \\
  \hline
  \end{tabular}}
\end{table}

\section{Methodology}

\paragraph{Data}
We use publicly available data from NCI Imaging Data Commons (IDC), a cloud-based repository of cancer imaging data \cite{Fedorov2023} as seen in Table~\ref{table1}. The ground truth series type was assigned semi-automatically by manually checking all possible regular \textit{SeriesDescription} expressions specific for these datasets, where 84.1\% of all series were assigned to the T2-weighted (T2W), diffusion-weighted imaging (DWI), apparent diffusion coefficient (ADC) and dynamic contrast enhanced (DCE) classes. Series not included consisted of those acquired sagitally/coronally, localizers, and calculated b-value DWI images. %Series not assigned consisted of calculated b value DWI images, localizers, or sagitally/coronally acquired images. 
We split the data patient-wise, using 60\% of QIN-Prostate-Repeatability \cite{Fedorov2018A} and ProstateX \cite{Litjens2014, Litjens2017} for training, and 20\% of the same datasets for validation. Both collections contain the sequences to classify, including T2W, DWI, ADC and DCE images. We split testing into internal (20\% of QIN-Prostate-Repeatability and ProstateX), and external (five collections not seen during training). These external collections were curated in the same manner, to only contain the four classes used for the classification.

% The ground truth scanning sequence was assigned based on derived regular expressions for the \textit{SeriesDescription}, where 84.1\% of all series were assigned in this manner. On a patient level, we use 60\% of QIN-Prostate-Repeatability \cite{Fedorov2018A} and 60\% of ProstateX \cite{Litjens2014, Litjens2017} for training, and for validation we use 20\% of the same datasets. Both collections contain the sequences to classify, including T2W, DWI, ADC and DCE images. For the internal test set we use the last 20\%, while the external test set consists of the other five collections, which each include a portion of the four scans used for classification. 

\paragraph{Methods}
We propose a CNN-based method that leverages image data and DICOM metadata. The following metadata attributes were used, as these are machine-generated, standardized, and not removed during de-identification: \textit{RepetitionTime}, \textit{EchoTime}, \textit{FlipAngle}, \textit{ScanningSequence}, and \textit{ContrastBolusAgent}. 
% The \textit{ScanningSequence} parameter was reformulated as three True/False parameters to describe the echo planar (EP), spin echo (SE) and gradient recall (GR) sequences. The \textit{ContrastBolusAgent} parameter was reformulated as True/False. 
A derived \textit{is4D} attribute was assigned based on whether spatially overlapping slices were detected within the series. Feature scaling was applied to relevant metadata. Figure~\ref{hiplot} summarizes the metadata distribution. Center slices were extracted from 3D volumes, resampled to 64x64, and normalized between 0 and 1. The CNN consisted of three convolution layers each followed by max pooling, followed by three dense layers, where the metadata was concatenated. Sparse categorical cross entropy loss was used with Adam optimization. K-fold cross validation was performed (4 folds), where each model was trained for 10 epochs with early stopping, and probability outputs were ensembled. The combined image and metadata approach was compared to A) a random forest classifier using only the metadata, and a B) CNN model utilizing only the images.

\begin{figure}[]
\floatconts
  {hiplot}
  {\caption{Hiplot visualization \cite{hiplot} of DICOM metadata parameters, colored by the assigned ground truth scan type.}}
  {\includegraphics[width=0.9\linewidth, trim={0cm 0cm 0cm 0cm},clip]{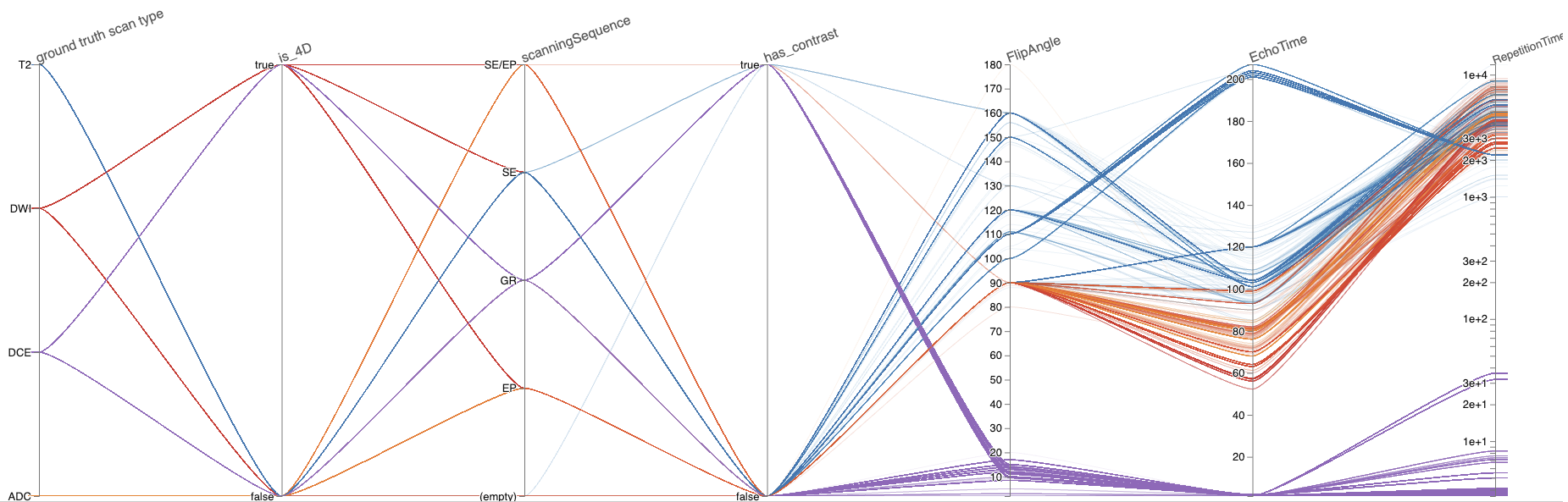}}
\end{figure}
%\vspace*{-20pt}
\begin{table}[]
\centering
  \tiny
  {\caption{Quantitative results for the three methods on the two test datasets. The mean is provided for the F-beta score when four fold cross-validation is performed. %the mean is provided for the F-beta score.
  }\label{table2}}%
  {\begin{tabular}{|c|c|c|c|c|c|c|c|c|}
  \hline
  \bfseries Method & \bfseries Internal & \bfseries Internal & \bfseries Internal & \bfseries  Internal & \bfseries  External & \bfseries  External & \bfseries External & \bfseries External \\
  & \bfseries T2Ax & \bfseries DWI & \bfseries ADC & \bfseries DCE & \bfseries T2Ax & \bfseries DWI & \bfseries ADC & \bfseries DCE \\
  \hline
  Metadata & 1.00 & 1.00 & 1.00 & 1.00 & 0.98 & 0.60 & 0.91 & 1.00\\
  Images & 0.99 & 0.99 & 0.99 & 0.99 & 0.93 & 0.59 & 0.99 & 0.89\\
  Images + metadata & 1.00 & 1.00 & 0.99 & 0.99 & 0.98 & 0.72 & 0.99 & 0.99\\
  \hline
  \end{tabular}}
\end{table}

%%%%%%%%%%%%%%%%%%%%%%%%%%%%%%%%%%%%%%%%%%%%%%%%%%%%%%%%%%%%%%%%%%%%%%%%%%%%%%%%%%%%%%%%%%%%%%%%%%

\section{Results and Discussion}
Table~\ref{table2} displays the evaluation. We note the higher accuracy on the internal dataset, as patients are from the same collection as training/validation, compared to the external test set, which contains five collections not seen during training. The metadata-only approach performs poorly on the external test set due to DWI misclassified as ADC and vice versa. Depending on the format of the DWI data (single series is a 4D volume, or multiple series each with a 3D volume), the latter can be considered similar to ADC. In the images and metadata combination approach, the performance of DWI in the external test improves, but DWI still suffers from misclassification as ADC due to the \textit{is4D} parameter, and the intensity similarity of the two. Future work involves refining the metadata used, and including the ability to differentiate between low and high b-value images.

\midlacknowledgments{We acknowledge Dr. Clare Tempany, grants P41EB028741-04, P41EB028741-03S1, and NIH NCI under Task Order No. HHSN2611 0071 under Contract No. HHSN261201500003l, which made this research possible.}

%%%%%%%%%%%%%%%%%%%%%%%%%%%%%%%%%%%%%%%%%%%%%%%%%%%%%%%%%%%%%%%%%%%%%%%%%%%%%%%%%%%%%%%%%%%%%%%%%%

\bibliography{bibliography}

%%%%%%%%%%%%%%%%%%%%%%%%%%%%%%%%%%%%%%%%%%%%%%%%%%%%%%%%%%%%%%%%%%%%%%%%%%%%%%%%%%%%%%%%%%%%%%%%%%

\end{document}